\documentclass[twocolumn,
              superscriptaddress,
              floatfix,
              longbibliography, 
              showkeys,apl]{revtex4-2}

\usepackage{graphicx} 
\usepackage{bm}	
\usepackage{color}                     
\usepackage[dvipsnames]{xcolor}
\usepackage{epsfig}
\usepackage{amsmath} 
\usepackage{amssymb} 
\usepackage{longtable} 
\usepackage{float}
\usepackage{mathtools}
\usepackage{xparse}
\usepackage{hyperref}
\usepackage{times}
\usepackage[normalem]{ulem}
\usepackage{dsfont}

\usepackage{algorithm}
\usepackage{algpseudocode}
\algrenewcommand\algorithmicrequire{\textbf{Input:}}
\algrenewcommand\algorithmicensure{\textbf{Output:}}
\usepackage{enumitem}

\usepackage{sidecap}

\sidecaptionvpos{figure}{t}

\newcommand{\maxbd}{{\chi_{\text{max}}}}
\newcommand{\chiL}{\chi_{\text{left}}}
\newcommand{\chiR}{\chi_{\text{right}}}
\newcommand{\ufour}{U$(4)$\ }

\newcommand{\unitary}[1]{U_{#1}}
\newcommand{\unitarydagger}[1]{U^\dagger_{#1}}
\newcommand{\layer}[1]{\text{L}[U]^{(#1)}}
\newcommand{\layerdagger}[1]{\text{L}[U]^{(#1)\dagger}}
\newcommand{\initstate}{0^{\otimes N}}
\newcommand{\bra}[1]{\langle #1 |}
\newcommand{\ket}[1]{| #1 \rangle}
\newcommand{\overlap}[2]{\langle #1 | #2 \rangle}

\newcommand{\protocolone}{\mbox{$D_{all}$}\ }
\newcommand{\protocoltwo}{\mbox{Iter$[O_i D_{i}]$}\ }
\newcommand{\protocolthree}{\mbox{$O_{all}$}\ }
\newcommand{\protocolfour}{\mbox{$D_{all} O_{all}$}\ }
\newcommand{\protocolfive}{\mbox{Iter$[I_i O_{all}]$}\ }
\newcommand{\protocolsix}{\mbox{Iter$[D_i O_{all}]$}\ }
\newcommand{\complexitymps}{\mathcal{C}_{\text{MPS}}}
\newcommand{\complexitydecomp}{\mathcal{C}_{\text{Decomp}}}
\newcommand{\complexityopt}{\mathcal{C}_{\text{Opt}}}

\begin{document}

\setlist[enumerate,1]{label=\arabic*, start=0}

\title{Decomposition of Matrix Product States into Shallow Quantum Circuits}

\author{Manuel S. Rudolph}
\affiliation{Zapata Computing Canada Inc., 325 Front St W, Toronto, ON, M5V 2Y1}

\author{Jing Chen}
\affiliation{Zapata Computing Inc., 100 Federal Street, Boston, MA 02110, USA}

\author{Jacob Miller}
\affiliation{Zapata Computing Canada Inc., 325 Front St W, Toronto, ON, M5V 2Y1}

\author{Atithi Acharya}
\affiliation{Zapata Computing Inc., 100 Federal Street, Boston, MA 02110, USA}
\affiliation{Rutgers University, 136 Frelinghuysen Rd, Piscataway, NJ 08854, USA}

\author{Alejandro Perdomo-Ortiz}
\email{alejandro@zapatacomputing.com}
\affiliation{Zapata Computing Canada Inc., 325 Front St W, Toronto, ON, M5V 2Y1}


\date{\today} 

\begin{abstract}
	\textit{Tensor networks} (TNs) are a family of computational methods built on graph-structured factorizations of large tensors, which have long represented state-of-the-art methods for the approximate simulation of complex quantum systems on classical computers. The rapid pace of recent advancements in numerical computation, notably the rise of GPU and TPU hardware accelerators, have allowed TN algorithms to scale to even larger quantum simulation problems, and to be employed more broadly for solving machine learning tasks. The ``quantum-inspired'' nature of TNs permits them to be mapped to \textit{parametrized quantum circuits} (PQCs), a fact which has inspired recent proposals for enhancing the performance of TN algorithms using near-term quantum devices, as well as enabling joint quantum-classical training frameworks which benefit from the distinct strengths of TN and PQC models. However, the success of any such methods depends on efficient and accurate methods for approximating TN states using realistic quantum circuits, something which remains an unresolved question.
	
	In this work, we compare a range of novel and previously-developed algorithmic protocols for decomposing \textit{matrix product states} (MPS) of arbitrary bond dimension into low-depth quantum circuits consisting of stacked linear layers of two-qubit unitaries. These protocols are formed from different combinations of a preexisting analytical decomposition scheme with constrained optimization of circuit unitaries, and all possess efficient classical runtimes. Our experimental results reveal one particular protocol, involving sequential growth and optimization of the quantum circuit, to outperform all other methods, with even greater benefits seen in the setting of limited computational resources. Given these promising results, we expect our proposed decomposition protocol to form a useful ingredient within any joint application of TNs and PQCs, in turn further unlocking the rich and complementary benefits of classical and quantum computation.
	
\end{abstract}

\maketitle

\section{Introduction}\label{sec:intro}

Parametrized quantum algorithms represent a new and promising technology for tackling a variety of problems in fundamental research and real-world applications alike. While the term ``quantum algorithm'' is commonly associated with quantum circuit algorithms running on quantum computers, there also exists a family of quantum (commonly referred to as ``quantum-inspired'') algorithms whose evaluation is classically tractable using \textit{tensor network} (TN) models~\cite{TNorg}. 
The field of TNs is developing at a fast pace, with TNs enabling state-of-the-art performance in quantum simulation~\cite{orus2019tensor, pan2022solving, zhou_2020}, and more recently also machine learning (see e.g.,~\cite{Huggins_2019, stoudenmire2016supervised, Novikov_2016, cohen2016expressive, han2018unsupervised, rudolph2022synergistic} and combinatorial optimization~\cite{alcazar2021enhancing, mugel2020dynamic, liu2021tropical}. In contrast to quantum algorithms on quantum computers, TNs allow for an analytical construction of the encoded quantum system on classical computers which enables powerful tools such as low-rank factorization, and efficient exact calculation of state amplitudes and fidelities. 
Additionally, the abundance and comparative cheapness of classical hardware, along with GPU and TPU support, has allowed TNs to handle simulations of extraordinary scale~\cite{pan2020contracting, gray2021hyper, Pan2022}. 
On the other hand, TN models are restricted in the amount of long-range entanglement that they can encode in a quantum state, which can be a limiting factor in some quantum simulation or machine learning applications~\cite{eisert2010colloquium, martyn2020entanglement, lu2021tensor}.

The counterpart to TNs on quantum computers are \textit{parametrized quantum circuits} (PQCs), which are currently one of the most promising frameworks for effectively utilizing near- and mid-term quantum computers with limited numbers of qubits and modest amounts of hardware noise~\cite{benedetti2019parameterized,Cerezo2021_vqareview,Bharti2022_nisqreview}. 
With their distinct strengths and weaknesses relative to TNs, PQCs promise to help solve some tasks more effectively than TNs by enabling long-range entangling operations to prepare quantum states~\cite{Aaronson-2015, Preskill2018, national2019quantum}. 

Owing to their similar mathematical structure and shared use for parametrizing quantum states, both TNs and PQCs can be transformed exactly or approximately into instances of the other~\cite{shi2006classical, markov2008simulating}. Transferring previously optimized quantum states from classical to quantum hardware may not only reduce the amount of time and resources spent on expensive and scarce quantum computers, but it may enable new state-of-the-art results with joint optimization frameworks.
Initial stages of the quantum state optimization could first be simulated using TNs, before continuing on quantum computers with access to increased model expressivity through long-range entangling gates (see e.g., Ref.~\cite{rudolph2022synergistic} for a concrete realization using the developments of this work). 
The success of such joint schemes then clearly hinges on the quality of the transfer from TNs to PQCs, as well as the depth of the resulting quantum circuit. The latter is a critical requirement for near-term quantum computers, whose noise may lead to significant decay in state fidelity for deep circuits.
While joint frameworks utilizing TNs and PQCs have already been proposed, and are expected to yield significant benefits~\cite{Huggins_2019}, there remains a relative scarcity of computational techniques for converting TN models to high-quality shallow quantum circuits, with best practices for this conversion process still lacking.


In this work, we introduce a number of algorithmic protocols for approximately decomposing \textit{matrix product states} (MPS), a one-dimensional flavor of TNs, of arbitrary bond dimension $\chi$ into stacked linear layers of two-qubit unitaries. These protocols each utilize different combinations of the analytic decomposition technique proposed in Ref.~\cite{ran2020encoding} with constrained optimization of circuit unitaries on classical hardware, and include as special cases the decomposition methods of Refs.~\cite{ran2020encoding, lin2021real}, the more sophisticated method of Ref.~\cite{dov2022approxencoding}, and layerwise PQC training strategies~\cite{skolik2021layerwise}. We benchmark these protocols on a variety of MPS arising from many-body and machine learning settings, and find one particular protocol, involving sequential growth and optimization of the quantum circuit, that emerges as the most successful technique. This novel decomposition method consistently delivers lower approximation errors than other protocols, with an advantage that becomes more pronounced with a limited computational budget. This decomposition protocol can be used as an effective tool for mapping MPS to shallow quantum circuits with high fidelity, which we anticipate will make possible the realization of powerful joint frameworks utilizing both TNs and PQCs.

\begin{figure*}
	\centering
	\includegraphics[width=0.8\linewidth]{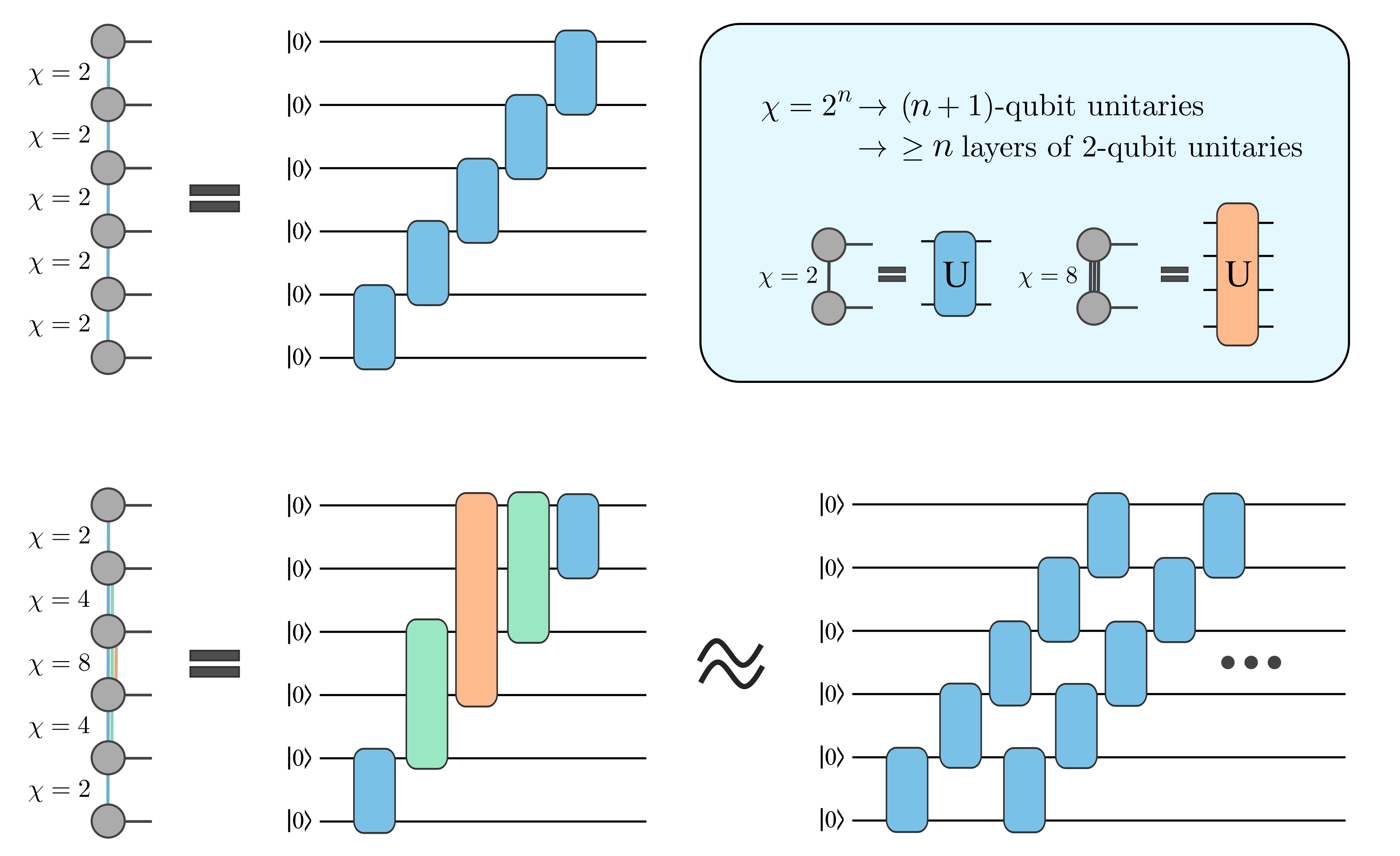}
	\caption{Exact or approximate mapping of an MPS with an arbitrary maximum bond dimension $\maxbd$ to a quantum circuit. (Top left) MPS with $\maxbd=2$ can be mapped exactly to a quantum circuit consisting of one linear layer of two-qubit unitaries. (Bottom, left) In general, for an MPS with a maximum bond dimensions of $\maxbd$, each bond in the MPS with bond dimension $\chi$ maps to a multi-qubit unitary acting on $\left \lceil\log_2( \chi)\right \rceil +1$ qubits. (Bottom right) To more effectively transport  MPS to quantum hardware, we study the decomposition of MPS into multiple layers of two-qubit unitaries. 
	}
	
	\label{fig:mps_mapping}
\end{figure*}

\section{Background}\label{sec:background}

\subsection{Tensor Networks}\label{ssec:tensor_networks}
Tensor networks are low-rank factorizations of large tensors into networks of smaller tensors, which can represent wave functions and quantum circuits on classical hardware~\cite{orus2014practical,TNorg}. TNs are typically illustrated as undirected graphs, whose nodes represent ``core'' tensors containing the parameters of the model, and whose edges indicate contractions between core tensors along particular tensor axes. TN model families are defined by their underlying graph structures, with 1D line graphs giving MPS~\cite{garcia_mps}, 2D grid graphs giving \textit{projected entangled pair states} (PEPS)~\cite{Verstraete2004}, and tree graphs (or equivalently, acyclic graphs) giving \textit{tree tensor networks} (TTNs)~\cite{shi2006classical}. The main capacity parameter of TNs is the dimension of the vector spaces being contracted over, known as the bond dimension $\chi$. This parameter controls the expressive power of the model, as well as the computational resources and memory requirements for classical simulations. The bond dimensions needed to encode a target wavefunction can be estimated by the entanglement entropy within the target state~\cite{vidal2003efficient}, with states possessing entanglement entropy $S$ requiring TNs of bond dimensions $\chi \sim e^S$ to exactly represent. MPS implementations typically utilize bond dimensions of at most $\chi \sim 10^3$, although high-performance simulations using MPS are able to reach bond dimensions of $\chi \sim 10^4 - 10^5$~\cite{ganahl2022density}. 

TN simulations of quantum circuits can be very effective when the graph structure of the TN matches the circuit layout~\cite{markov2008simulating, eisert2010colloquium}, with the \textit{singular value decomposition} (SVD) enabling efficient approximate simulations by dropping lower magnitude singular values in the Schmidt decomposition of the target state~\cite{vidal2003efficient}. As a concrete example, TN simulations in Ref.~\cite{pan2022solving} were able to achieve higher overall state fidelity when simulating the multi-layer 2D circuits of the so-called \textit{quantum supremacy} experiments in Ref.~\cite{Google2019supremacy}, where the fidelity deteriorated due to quantum hardware errors. On the other hand, computational cost of classical simulations can quickly escalate when the TN graph and quantum circuit layouts differ, such that circuit gates act on qubits that are far apart on the TN graph. In such cases, the bond dimensions necessary for a high-fidelity simulation will generally increase exponentially with the system size, owing to the need to carry entanglement between distant qubits through all intermediate edges.

\subsection{Canonical forms and Isometries}\label{ssec:background_isometries}
While simulating quantum circuits on classical hardware via representations as TNs has many fundamental use-cases, our focus here is on the reverse direction, of efficiently representing TNs as the output of quantum circuits. An important tool for carrying out this latter conversion are TN \textit{canonical forms}, which facilitate the loss-less conversion of all core tensors in a TN into isometries (i.e. inner product preserving transformations between Hilbert spaces). While weaker canonical forms have been developed for general TNs~\cite{evenbly2018gauge}, the ability to efficiently convert core tensors into isometries requires TN architectures to be defined on acyclic graphs, such as MPS and TTNs.

By restricting to bond dimensions which are powers of 2, the isometries arising in a TN in canonical form are equivalent to matrices of shape $2^n\times 2^m$ (assuming $n \geq m$ without loss of generality), which can always be expanded to $2^n \times 2^n$ unitary matrices (i.e. $n$-qubit unitaries). 
In more detail, if $Q$ is an isometry of shape $2^n\times 2^m$ satisfying $Q^{\dagger}Q=\mathds{1}_{2^m \times 2^m}$, 
the construction of the corresponding $n$-qubit unitary $U$ involves extending the $2^m$ columns of $Q$ with $\kappa := 2^n - 2^m$ basis vectors from the kernel space of $Q^\dagger$. These additional basis vectors can be arranged in a $2^n \times \kappa$ matrix $X$ satisfying $Q^{\dagger}X=0$ and $X^\dagger X = \mathds{1}_{\kappa \times \kappa}$, which guarantees that the $2^n\times 2^n$ square matrix $U = [Q \quad X]$ satisfies the unitary condition $U^{\dagger}U=U U^\dagger = \mathds{1}_{2^n \times 2^n}$. 

For a general MPS parametrizing an $N$-qubit wavefunction, each tensor core will have left and right bond dimensions of $\chiL$ and $\chiR$, which, by padding with zeros, can be expanded to the nearest powers-of-two, namely $\chiL' = 2^n$ and $\chiR' = 2^m$, where $n = \lceil\log_2(\chiL)\rceil$, $m = \lceil\log_2(\chiR)\rceil$, and $\lceil \cdot \rceil$ denotes the ceiling operation. Placing this MPS in (left) canonical form via iterated QR decompositions (see Ref.~\cite{garcia_mps} for details) will then yield a $2^{n+1} \times 2^m$ isometry, which can then be converted into a unitary gate acting on $n+1$ qubits.
The linear topology of the MPS leads the composition of gates from this conversion process to form one layer of multi-qubit unitaries arranged in a staircase topology, which we refer to as a linear circuit layer.
Fig.~\ref{fig:mps_mapping} illustrates this exact conversion from MPS to PQC for an MPS with $\maxbd = 8$.

Despite the simplicity of the conversion process described above, a major deficit of this procedure is the fact that arbitrary $n+1$ qubit unitary gates cannot be directly implemented in quantum circuits running on real quantum hardware, which typically only support gates acting on one or two qubits. Overcoming this limitation while still obtaining an efficient and accurate conversion from TNs to PQCs is a primary motivation for our work.

\subsection{Related Work}\label{sec:related}

The problem of efficiently representing general MPS as the output of PQCs is crucial for many applications in condensed matter physics and quantum machine learning (QML) which utilize both classical and quantum computational resources. In the former domain, Refs.~\cite{lubasch2020variational, lin2021real, barratt2021parallel, smith2022crossing} explored the use of PQCs as quantum variational ansätze for calculating diverse properties of complex many-body systems, and identified distinct benefits of this method relative to the use of TNs on their own. Within QML, Ref.~\cite{Huggins_2019} introduced the idea of a joint optimization framework utilizing both TNs and PQCs, which the authors predicted would give improved performance in QML tasks. While this proposal lacked a concrete method for transferring TNs solutions to quantum computers, the PQC pretraining method of Ref.~\cite{dborin2022pretraining} gave numerical verification of the benefits of this style of joint training for MPS with $\maxbd=2$ (which exactly decompose into a single linear layer of two-qubit unitaries). The recent work of Ref.~\cite{rudolph2022synergistic}, which utilizes the conversion method described here, illustrates the increased boost in QML task performance that results from the use of larger TN models for initializing PQCs.

At the level of specific algorithms, Ref.~\cite{ran2020encoding} proposed an analytic technique for decomposing MPS of arbitrary $\maxbd$ into multiple linear layers of two-qubit unitaries, where the unitaries for each layer are chosen based on truncations of the MPS to $\chi=2$ across each bond, determined through iterated SVDs. This has the advantage of requiring relatively small computational resources, but unfortunately leads to relatively small gains with increasing circuit layers, decreasing fidelity past a critical number of layers, and poor performance compared to more sophisticated approaches. In a different direction, Ref.~\cite{lin2021real} proposed decomposing arbitrary MPS into circuits of stacked linear layers via iterative optimization over the two-qubit unitaries forming that circuit. This optimization method is similar to the Evenbly-Vidal algorithm~\cite{evenbly2009algorithms}, as well as the procedure described in Sec.~\ref{ssec:optimization}, and can support more general circuit architectures and achieve better performance via greater numbers of optimization sweeps. However, without a high-quality initialization for the circuit gates, this method remains vulnerable to barren plateaus~\cite{Mcclear2018Barren}, and becomes expensive when large numbers of sweeps are needed.

Closer to our work is the recent Ref.~\cite{dov2022approxencoding}, which utilizes both analytical and optimization-based decomposition methods to efficiently approximate arbitrary MPS using PQCs. This method first initializes the gates of a circuit using the analytical method, before using iterated optimization of the unitary gates to improve fidelity with the target MPS, a combination that was shown to have significant performance benefits over that of Ref.~\cite{ran2020encoding}. This method is only one of many decomposition protocols we benchmark here (denoted by \protocolfour in the language of Sec.~\ref{ssec:combination_decomposition}), and while we confirm the clear advantages of combining analytical and optimization methods, our empirical findings show that there are yet better ways of utilizing these two styles of decompositions to approximate arbitrary MPS using quantum circuits.

\section{Methods}\label{sec:methods}

In this Section, we outline the fundamental algorithmic building blocks used throughout our work, as well as the MPS decomposition protocols that can be built from them. To resolve any possible ambiguity that might arise, we give a detailed description of our notation and indexing convention for unitaries $U$ and layers of unitaries L$[U]$ in Appendix~\ref{apx:conventions}.

\subsection{Analytic Decomposition by Disentangling}\label{ssec:analytic}
\begin{figure*}[t]
	\centering
	\includegraphics[width=0.95\linewidth]{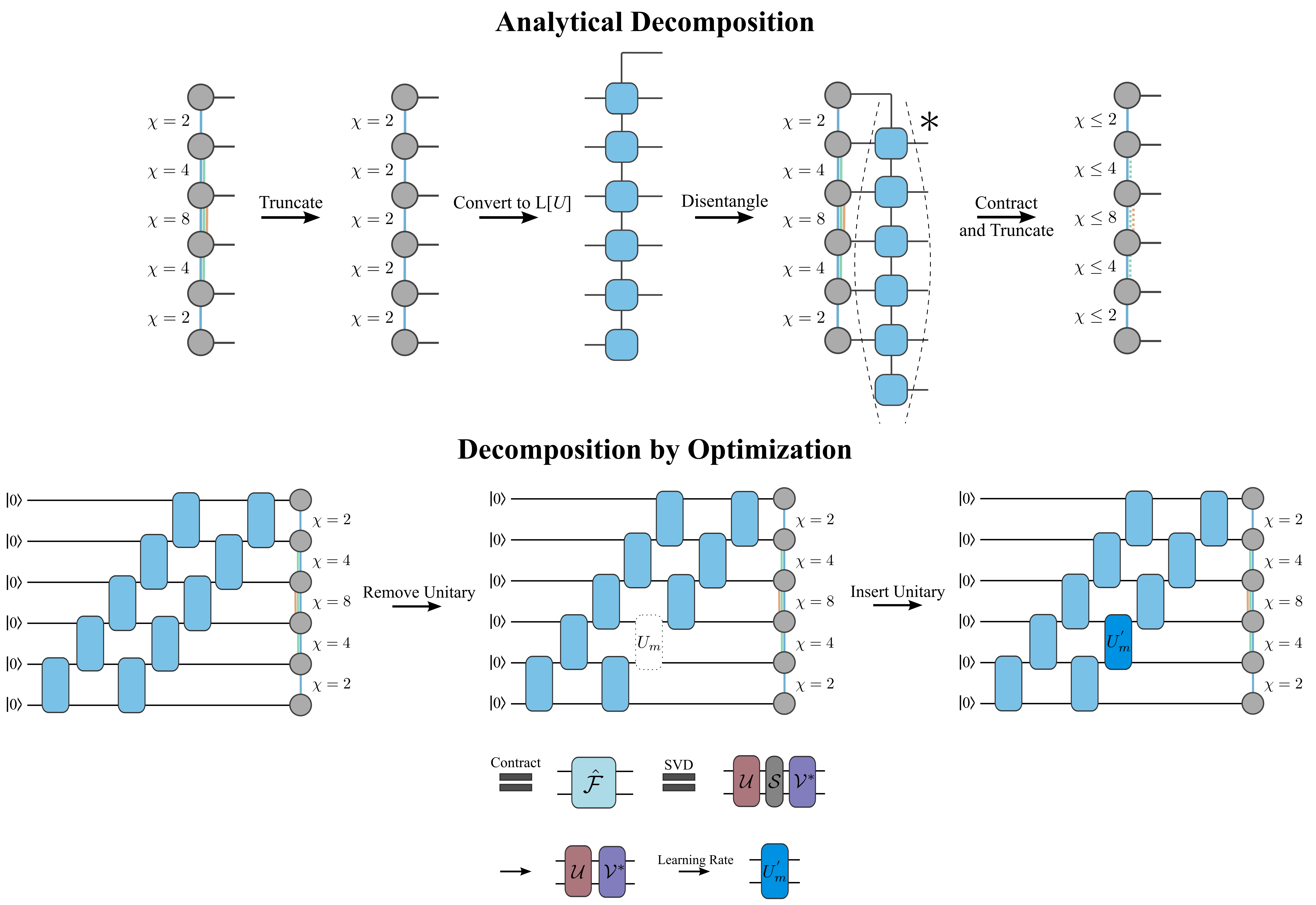}
	\caption{Illustration of the two fundamental building-block algorithms that are used to compose practical decomposition protocols for an MPS with arbitrary bond dimension $\maxbd$ into linear layers of two-qubit unitaries. (TOP) Analytical decomposition algorithm presented in Ref.~\cite{ran2020encoding}. See Sec.~\ref{ssec:analytic} for details. A $\maxbd=2$ MPS map exactly to one layer of two-qubit unitaries (see also Fig.~\ref{fig:mps_mapping}), and one can truncate an MPS via SVD to $\chi=2$ such that the fidelity $|\overlap{\psi_{\chi=2}}{\psi_\maxbd}|$ is near optimal. The $\chi=2$ version of the MPS maps to one layer L$[U]$ of two-qubit unitaries, whose inverse can then be applied to disentangle the MPS. The process can be repeated iteratively to create several unitary layers that approximate the original MPS state. (Bottom) A constrained optimization algorithm for unitaries in TNs which is demonstrated in Ref.~\cite{shirakawa2021automatic}. See Sec.~\ref{ssec:optimization} for details. Updates for a circuit unitary $\unitary{m}$ are calculate by first calculating the environment tensor $\hat{\mathcal{F}}_m$ in Eq.~\eqref{eq:environment_tensor}, and applying SVD on it to find the locally optimal unitary. A learning rate is applied to improve the convergence behavior.
	}
	\label{fig:fundamental_algorithms}
\end{figure*}

Ref.~\cite{ran2020encoding} presents a technique to sequentially decompose an MPS into $K$ layers of two-qubit unitaries with a linear next-neighbor topology. The approach is aimed at maximizing the fidelity 
\begin{equation}\label{eq:fidelity_analytic}
	\begin{aligned}
		f\big(\{\layer{k}\}_{k=1}^K\big) 
		&= \big|\bra{\initstate}\prod_{k=1}^K \layerdagger{k}\ket{\psi_{\maxbd}}\big|\\
		&= \big|\bra{\initstate}\layerdagger{K}\dots\layerdagger{1}\ket{\psi_{\maxbd}}\big|\\
		&= \big|\overlap{\psi_{QC}^{(K)}}{\psi_{\maxbd}}\big|
	\end{aligned}
\end{equation}
between the original MPS state $\ket{\psi_{\maxbd}}$, and a quantum state 
\begin{equation}
	\begin{aligned}
		\ket{\psi_{QC}^{(K)}} 
		&=  \prod_{k=K}^1 \layer{k}\ket{\initstate}\\
		&=  \layer{1}\dots\layer{K}\ket{\initstate}
	\end{aligned}
\end{equation}
which is meant to approximate the MPS. The resulting quantum circuit is $\prod_{k=K}^1 \layer{k}$, and it consists only of two-qubit \ufour unitaries. Equivalently, one could view the inverse layers $\prod_{k=1}^K \layerdagger{k}$ as the circuit that \textit{disentangles} the MPS.

Algorithm~\ref{alg:analytic} describes the analytic decomposition technique, which which is also illustrated in Fig.~\ref{fig:fundamental_algorithms} (top).
\begin{algorithm}[H]
	\caption{Analytic Decomposition}\label{alg:analytic}
	\begin{algorithmic}[1]
		\Require MPS $\psi_{\maxbd}$, Maximum layers $K$, Target fidelity $\hat{f}$ 
		\Ensure Quantum Circuit layers $\prod_{k=K}^1 \layer{k}$
		\State $k \gets 0$
		\State $\ket{\psi^{(0)}}  \gets \ket{\psi_{\maxbd}}$
		\While{$k<K$ or $|\overlap{\initstate}{\psi^{(k)}}| < \hat{f}$}
		\State Truncate $\psi^{(k)}$ to $\psi^{(k)}_{\chi=2}$ via SVD
		\State Convert $\psi^{(k)}_{\chi=2}$ to $\layer{k+1}$
		\State $|\psi^{(k+1)}\rangle \gets \layerdagger{k+1} |\psi^{(k)}\rangle$
		\State $k \gets k+1$
		\EndWhile
	\end{algorithmic}
\end{algorithm}
In this algorithm, one truncates a copy of the original MPS $\psi_{\maxbd}$ to $\chi=2$ via SVD, converts the truncated MPS $\psi_{\chi=2}$ to one layer L$[U]$ of two-qubit unitaries, and applies the inverse of that layer L$[U]^\dagger$ to the original MPS. The resulting MPS is expected to have less entanglement and the bond dimensions may decrease depending on the predefined singular value thresholds. The conversion of $\chi=2$ MPS to a linear layer of two-qubit unitaries is explained in Sec.~\ref{ssec:background_isometries} and Refs.~\cite{ran2020encoding,barratt2021parallel,dborin2022pretraining}. This process can be repeated iteratively to create several layers $\prod_{k=K}^1 \layer{k}$ which are indexed from $K$ to $1$ because the layers are added ``from the MPS backwards''. Consequently, the newest layer $K$ is at the beginning of the resulting quantum circuit.

The analytical decomposition algorithm is relatively inexpensive to perform, and the decomposition can be improved sequentially up until a desired fidelity, with every additional layer likely giving better decomposition performance. 
However, it is crucial to note that for a fixed number of layers $K$, it is not natively possible to increase the quality of the decomposition. This leads to our observation that $K$ sometimes needs to be orders of magnitudes larger for high-fidelity decomposition than predicted by the lower bound \mbox{$K_{min} \sim \log_2(\maxbd)$}~\cite{schuch2008entropy}. For this reason, we believe that this algorithm is best utilized in conjunction with an effective constrained optimization algorithm for the resulting quantum circuit unitaries.

\subsection{Decomposition by Optimization}\label{ssec:optimization}
We now describe the constrained optimization algorithm for the circuit unitaries used throughout this work. It was demonstrated in Ref.~\cite{shirakawa2021automatic} for learning quantum circuits to prepare quantum states.
Given an initial quantum circuit $\prod_{m=1}^MU_m$ consisting of $M$ unitaries $U_m$, the goal of the optimization algorithm is to increase the fidelity
\begin{equation}\label{eq:fidelity_optimization}
	\begin{aligned}
		f\big(\{U_m\}_m\big) 
		&= \big|\langle 0^{\otimes N}|\prod_{m=M}^1 U^\dagger_m|\psi_{\maxbd}\rangle\big|\\
		&= \big|\bra{\initstate}\unitarydagger{1}\dots\unitarydagger{M}\ket{\psi_{\maxbd}}\big|\\
		&= \big|\overlap{\psi_{QC}^{(M)}}{\psi_{\maxbd}}\big|
	\end{aligned}
\end{equation}
between the original MPS state $\ket{\psi_{\maxbd}}$, and a quantum state 
\begin{equation}
	\begin{aligned}
		\ket{\psi_{QC}^{(M)}} 
		&=  \prod_{m=1}^M \unitary{m}\ket{\initstate}\\
		&=  \unitary{M}\dots\unitary{1}\ket{\initstate}
	\end{aligned}
\end{equation}
which is meant to approximate the MPS. In our case, the unitaries $\unitary{m}$ act only on two qubits, i.e., the unitaries are \ufour matrices, and the circuit layout is a linear next-neighbor topology for efficient classical simulation using MPS, as well as for compatibility with Algorithm~\ref{alg:analytic} in Sec.~\ref{ssec:analytic}.

Algorithm~\ref{alg:optimization} describes the optimization algorithm used in this work, which is also illustrated in Fig.~\ref{fig:fundamental_algorithms} (bottom).
\begin{algorithm}[H]
	\caption{Decomposition by Optimization}\label{alg:optimization}
	\begin{algorithmic}[1]
		\Require Initial quantum circuit $\prod_{m=1}^M \unitary{m, 0}$, number of sweeps $T$, target fidelity $\hat{f}$, learning rate $r \in [0, 1]$
		\Ensure Optimized quantum circuit $\prod_{m=1}^M U_m$
		\State $t \gets 1$
		\While{$t<T$ or $\left | \langle 0^{\otimes N}| \prod_{m=M}^1 U^\dagger_m |\psi_{\chi_{max}}\rangle \right | < \hat{f}$}
		\For{$m$ in 1 \dots $M$}
		\State Calculate $\hat{\mathcal{F}}_m$ in Eq.\eqref{eq:environment_tensor}
		\State SVD $\hat{\mathcal{F}} = \mathcal{U}\mathcal{S}\mathcal{V^{\dagger}}$
		\State $ \unitary{\text{new},m} \gets \mathcal{U}\mathcal{V^{\dagger}}$ 
		\State Apply learning rate in Eq.~\eqref{eq:learning_rate}\par
		\hskip\algorithmicindent $\unitary{m} \gets \unitary{m}(\unitarydagger{m} \unitary{\text{new},m})^r$
		\EndFor
		\State $t \gets t+1$
		\EndWhile
	\end{algorithmic}
\end{algorithm}
The optimization algorithm iterates over all unitaries in the circuit and updates them one by one. The number of sweeps can either be predefined through a maximum number $T$, or until a target fidelity $\hat{f}$ is reached with $f\big(\{U_m\}_m\big) >  \hat{f}$ (compare Eq.~\eqref{eq:fidelity_optimization}). The update for one unitary $\unitary{m}$ can be calculated by first calculating the so-called \textit{environment} tensor for $U_m$:
\begin{equation}\label{eq:environment_tensor}
	\hat{\mathcal{F}}_m = \text{Tr}_{\Bar{U}_m}\left[\prod_{i=M}^{m+1} \unitary{i}\ket{\psi_{\maxbd}}\bra{\initstate}\prod_{j=1}^{m-1} \unitarydagger{j} \right].
\end{equation}
Here, the expression $\text{Tr}_{\Bar{U}_m}[\dots]$ denotes the partial trace over the qubit indices that don't interact with $U_m$. The environment tensor $\hat{\mathcal{F}}_m$ is represented as a $4\times 4$ matrix, and can in practice be calculated by ``removing'' $\unitary{m}$ from the circuit and contracting the remaining tensor network (see bottom of Fig.~\ref{fig:fundamental_algorithms}) while retaining the MPS structure throughout. Interestingly, $\hat{\mathcal{F}}_m$ is the operator which is locally optimal for the circuit to increase the fidelity with the MPS if it were replacing $\unitary{m}$~\cite{shirakawa2021automatic}, but this operator is not generally unitary. In order to keep all quantum circuit operations unitary, we can compute the SVD $\hat{\mathcal{F}}_m = \mathcal{U}\mathcal{S}\mathcal{V}^{\dagger}$, then use the fact that the product $\mathcal{U}\mathcal{V}^\dagger$ is the unitary matrix that minimizes the Frobenius distance with $\hat{\mathcal{F}}_m$~\cite{higham1988matrix}, which in turn maximizes the fidelity of the PQC with the MPS among all unitaries. This leads $\unitary{\text{new}} = \mathcal{U}\mathcal{V}^{\dagger}$ to be the locally optimal unitary for increasing the fidelity with the MPS.
Since this is a rather strong local update, we introduce a learning rate $r\in[0, 1]$, which influences the unitary update rule via
\begin{equation}\label{eq:learning_rate}
	\unitary{m}^{'} = \unitary{m}\left( \unitarydagger{m} \unitary{\text{new}} \right)^r.
\end{equation} 
We find the optimization process to give excellent performance using a value of $r=0.6$. We note that the fractional matrix exponential in $\left( \unitarydagger{m} \unitary{\text{new}} \right)^r$ can be implemented by simply applying the exponent to the diagonal elements in the eigendecomposition of $\unitarydagger{m} \unitary{\text{new}}$. 
Finally, $\unitary{m}^{'}$ is the unitary that replaces $\unitary{m}$, which gives improved fidelity with the MPS. 

We note that it is vital to the computational feasibility of this algorithm to perform optimization sweeps consisting of consecutive values for the index $m$,  because in that case the states used to compute $\hat{\mathcal{F}}_m$, i.e., $\prod_{i=M}^{m+1} \unitary{i}\ket{\psi_{\chi_{max}}}$ and $\bra{\initstate}\prod_{j=1}^{m-1} \unitarydagger{j}$, do not require full re-calculation.  With a proper choice of the optimization order, most of these states can be cached and reused. For example, after $\unitary{m}^{'}$ has been obtained, we apply $\unitary{m}^{'\dagger}$ to $\bra{\initstate}\prod_{j=1}^{m-1}$ and $\unitarydagger{m+1}$ to $\prod_{i=M}^{m+1} \unitary{i}\ket{\psi_{\chi_{max}}}$, which forwards the two sides of the TN to step $m+1$. While one can analogously perform backward sweeps, we restrict ourselves to forward sweeps in this work.

The issue with decomposing an MPS with this optimization algorithm alone (what we call the \textit{brute-force} method), is that when the initial circuit unitaries are random, the application of the circuit to the fully-optimized MPS state $\ket{\psi_{\maxbd}}$ can potentially increase the maximum bond dimension $\maxbd$ required to approximate the resulting quantum state without significant truncation of singular values. 
Therefore, we now combine Algorithms~\ref{alg:analytic} \&~\ref{alg:optimization}. Let Algorithm~\ref{alg:analytic} provide the initial circuit guesses that tend to \textit{decrease} the bond dimension of the MPS, and let Algorithm~\ref{alg:optimization} then optimize the unitaries to achieve the best approximation of the MPS for a fixed number of unitaries and layers.

\subsection{Combinations of analytic decomposition and optimization}\label{ssec:combination_decomposition}

\begin{figure}
	\centering
	\includegraphics[width=0.90\linewidth]{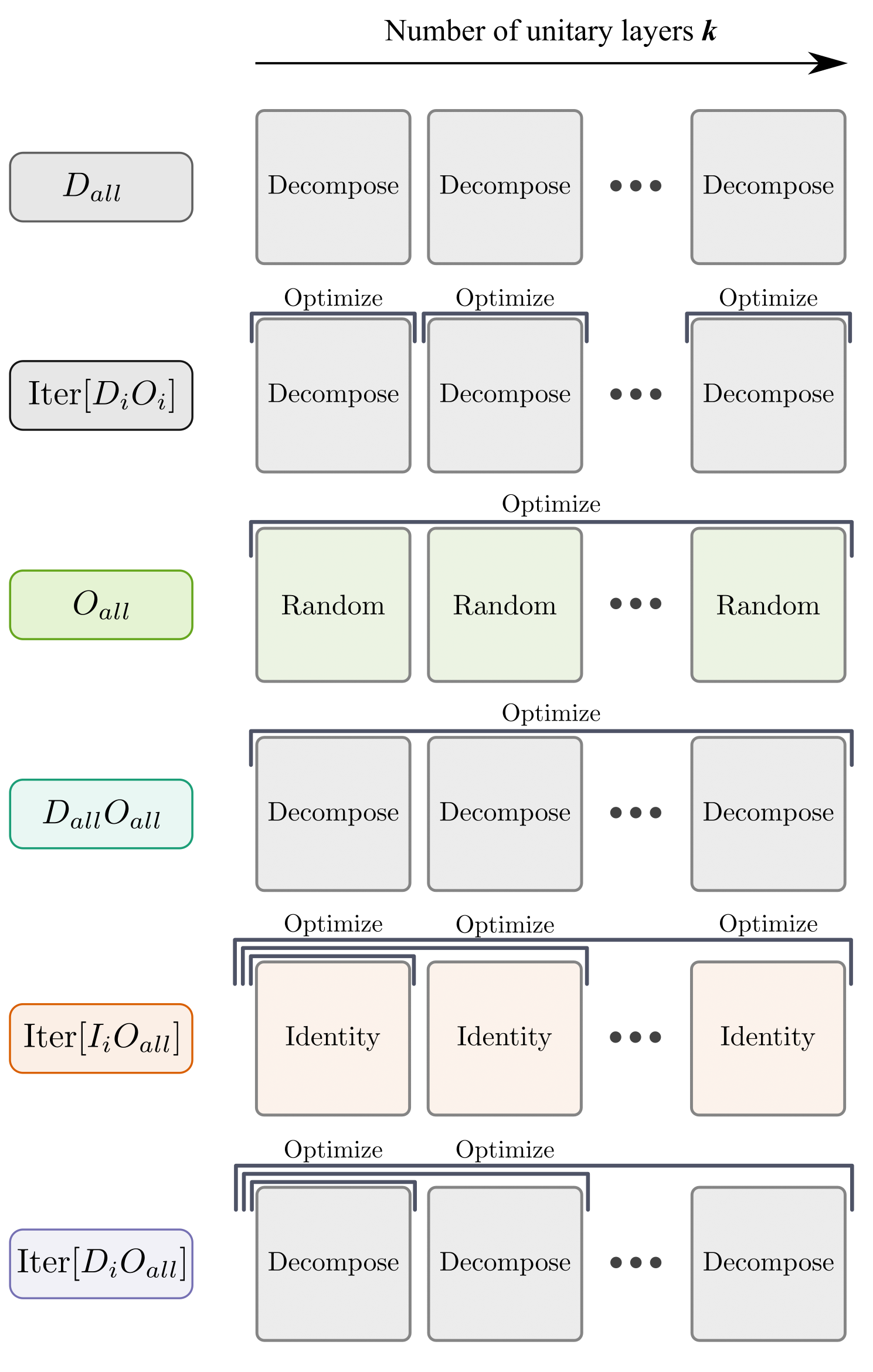}
	\caption{Schematic depiction of the MPS decomposition protocols studied in this work. The protocols utilize Alg.~\ref{alg:analytic} in Sec.~\ref{ssec:analytic} (as indicated by ``$D$''), as well as Alg.~\ref{alg:optimization} in Sec.~\ref{ssec:optimization} (as indicated by ``$O$''). The different protocols are described in Sec.~\ref{ssec:combination_decomposition}.}
	\label{fig:decomposition_protocols}
\end{figure}

In Secs.~\ref{ssec:analytic} and~\ref{ssec:optimization} we have introduced the fundamental algorithmic building blocks for decomposing MPS into quantum circuits consisting of only two-qubit unitaries. All steps of both algorithms are to be performed on classical hardware. Both approaches on their own have severe limitations, but complement each other very nicely.
To briefly summarize, while relatively inexpensive to perform, the main disadvantage of the analytic decomposition in Sec.\ref{ssec:analytic} is that the number $K$ of resulting quantum circuit layers can be very large to achieve a desired fidelity, and there is no way to increase fidelity for a fixed number of layers. Disadvantages of the brute-force decomposition algorithm in Sec.~\ref{ssec:optimization} include that random initial guesses for the unitaries may not only require many sweeps $T$ to optimize, but the random circuit can cause the bond dimension $\maxbd$ of the MPS to increase. However, optimization is a very valuable approach, because one spends classical resources to increase the TN decomposition quality without increasing circuit depth and potentially hampering the performance of the quantum circuits on noisy quantum hardware.

Fortunately, both algorithms can be combined into several different decomposition protocols, towards achieving the best trade-off between the required classical computing resources and the resulting quantum circuit depth. As quantum hardware improves, one can more reliably implement deeper circuits, which could in turn alleviate some of the burden of the MPS decomposition from classical hardware.

In Fig.~\ref{fig:decomposition_protocols}, we illustrate the combinations of Algorithms~\ref{alg:analytic} and~\ref{alg:optimization} that are studied in this work. The letter $D$ stands for the analytical decomposition algorithm, i.e., Algorithm~\ref{alg:analytic} in Sec.~\ref{ssec:analytic}. We make use of the possibility to either decompose one layer (indicated by $i$), or decompose to the full circuit depth $K$ (indicated by $all$). The letter $I$ represents an alternative approach to $D$, where linear layers are added to the quantum circuit in the form of identity unitaries. The letter $O$ stands for the application of the optimization algorithm, i.e., Algorithm~\ref{alg:optimization} in Sec.~\ref{ssec:optimization}. We either optimize the last layer that was created (indicated by $i$), or all layers in the circuit \textit{so far} (indicated by $all$). The ``Iter'' notation implies that there is an iterative procedure between creating new layers (either with $D$ or $I$), and optimizing the newest layer or all layers so far (indicated by $i$ or $all$, respectively).

Our first protocol is $D_{all}$, which is equivalent to applying only Algorithm~\ref{alg:analytic}. The second protocol is Iter$[D_iO_i]$, and it represents a minimal extension of $D_{all}$ where only the unitaries in the layer \mbox{$\layer{k+1}, k<K$} are optimized before the disentangling step \mbox{$\ket{\psi^{k+1}} \gets \layerdagger{k+1}\ket{\psi^{(k)}}$}. This aims to ensure that $|\overlap{\psi^{(k)}_{\chi=2}}{\psi^{(k)}}|$ is maximal in every iteration before disentangling, which SVD only ensures approximately. 
The third protocol is $O_{all}$, which is equivalent to applying only Algorithm~\ref{alg:optimization}. The circuit unitaries are initialized as $4 \times 4$ matrices with random entries from a normal distribution with zero mean. These are then converted into unitaries by selecting the unitary component $Q$ in the $QR$-decomposition of the random matrix.
The fourth protocol, $D_{all}O_{all}$, constitutes the first practical combination of both analytical decomposition and optimization. In this protocol, the $K$ circuit layers are first created using Algorithm~\ref{alg:analytic}, and then all unitaries are optimized in sweeps in further increase fidelity. The analytically decomposed layers provide a good starting point for the optimization algorithm as compared to random unitaries, and, crucially, the layers tend to already have a disentangling effect on the MPS, which reduces the bond dimension of the MPS. 
The fifth protocol, Iter$[I_iO_{all}]$, is a sequential protocol where new layers are iteratively added as identity operations, and and all existing layers are optimized using Algorithm~\ref{alg:optimization}. The sixth and final protocol studied in this work is Iter$[D_iO_{all}]$, where new layers are added via one analytical decomposition step, and all existing layers are optimized using Algorithm~\ref{alg:optimization}. The difference between the protocols Iter$[D_iO_i]$ and Iter$[D_iO_{all}]$ is that, while both maximize $|\overlap{\psi^{(k)}}{\psi_{\maxbd}}|$ at every decomposition step $k$, Iter$[D_iO_{i}]$ keeps layers $1\dots k-1$ fixed, however, these layers are also optimized in Iter$[D_iO_{all}]$.

The decomposition protocols studied in this work are constructed such that the protocol Iter$[D_iO_{all}]$ represents the most sophisticated combination of Algorithms~\ref{alg:analytic} \&~\ref{alg:optimization}. The other protocols can be seen as reference benchmarks with different components of Iter$[D_iO_{all}]$ being stripped away - either the analytical decomposition, the optimization, or the sequential growing scheme of the circuit. As such, we compare this protocol to previous work, where $D_{all}$ was proposed in Ref.~\cite{ran2020encoding}, $O_{all}$ is similar to the algorithm proposed in Ref.~\cite{lin2021real}, and $D_{all}O_{all}$ is similar to the method implemented in Ref.~\cite{dov2022approxencoding}.

\subsection{Discussion of Computational Complexity}\label{ssec:complexity}
It is crucial to the decomposition protocols presented in this work that all necessary steps can feasibly be performed on classical hardware. A rigorous derivation of the computational resources required for a practical decomposition of an MPS of interest is not within the scope of this work. Especially, because we don't aim to exactly represent an MPS, but we allow ourselves a certain error in the approximation. Therefore, in this section, we merely present a general overview of the expected complexities of Algorithms~\ref{alg:analytic} and~~\ref{alg:optimization}, as well as the protocols presented in Sec.~\ref{ssec:combination_decomposition}.

For an MPS with a maximal bond dimension $\maxbd$, the memory requirements of the MPS scale like $\sim N\chi^2_{\text{max}}$~\cite{orus2014practical}. 
However, including the cost of bringing the MPS into canonical form, the computational complexity $\complexitymps$ is
\begin{equation}
	\complexitymps \sim N\chi^3_{\text{max}}.
\end{equation}

Now approaching our MPS decomposition protocols, the protocol \protocolfour, for example, requires $K$ steps of analytical decomposition (see Sec.~\ref{ssec:analytic}), whose complexity $\complexitydecomp$ scales like
\begin{equation}
	\begin{aligned}\label{eq:complexity_decomp_plain}
		\complexitydecomp 
		&\sim K \cdot \complexitymps\\
		&\sim K N \chi^3_{\text{max}}.
	\end{aligned}
\end{equation}
We note that each linear layer has $N-1$ two-qubit unitaries, but the computational cost $\complexitymps$ already includes performing SVD on all bonds of the MPS. Then performing the $T$ optimization sweeps over all $K$ layers requires results in the complexity $\complexityopt$ of the optimization
\begin{equation}\label{eq:complexity_opt_plain}
	\begin{aligned}
		\complexityopt 
		&\sim K T \cdot \complexitymps\\
		&\sim N K T \chi^3_{\text{max}}.
	\end{aligned}
\end{equation}
The sequential protocols \protocoltwo, \protocolfive and \protocolsix apply $T$ optimization sweeps not only on $K$ layers, but on $k<K$ layers for all previous iterations $k$. However, these previous layers have fewer unitaries. The total number of layers that are optimized in a protocol that concludes at $K$ layers is $\frac{K(K+1)}{2}$. Thus, the sequential protocols (indicated by ``Iter'') receive an additional scaling factor of $K$:
\begin{equation}\label{eq:complexity_opt_sequential}
	\begin{aligned}
		\complexityopt^{\text{Iter}} \sim N K^2 T \chi^3_{\text{max}}.
	\end{aligned}
\end{equation}

The question now becomes, how many layers $K$ are needed for a high-fidelity approximation of the MPS? A convenient property of our main proposed protocol \protocolsix is that the circuit can grow sequentially, and one does not have to guess the circuit depth that can feasibly be optimized on classical hardware. However, a lower-bound of $K_{\text{min}} \sim \log_2(\maxbd)$ is presented in Ref.~\cite{schuch2008entropy}. We note that this bound provides a best-case circuit depth to exactly decompose a generic MPS, but exact decomposition is likely not the goal of a practical decomposition protocol. Therefore, we merely understand the scaling to provide a general sense of how deep quantum circuits consisting of two-qubit unitaries need to be to represent an MPS with maximum bond dimension $\maxbd$. In this case, we arrive at scalings of 
\begin{equation}\label{eq:complexity_scalings}
	\begin{aligned}
		&\complexitydecomp \sim N \chi^3_{\text{max}} \log_2(\maxbd)\\
		&\complexityopt \sim N T \chi^3_{\text{max}} \log_2(\maxbd) \\
		&\complexityopt^{\text{Iter}} \sim N T \chi^3_{\text{max}} \log_2(\maxbd)^2.
	\end{aligned}
\end{equation}

\section{Empirical Assessment}

\begin{figure*}[t]
	\centering
	\includegraphics[width=0.95\linewidth]{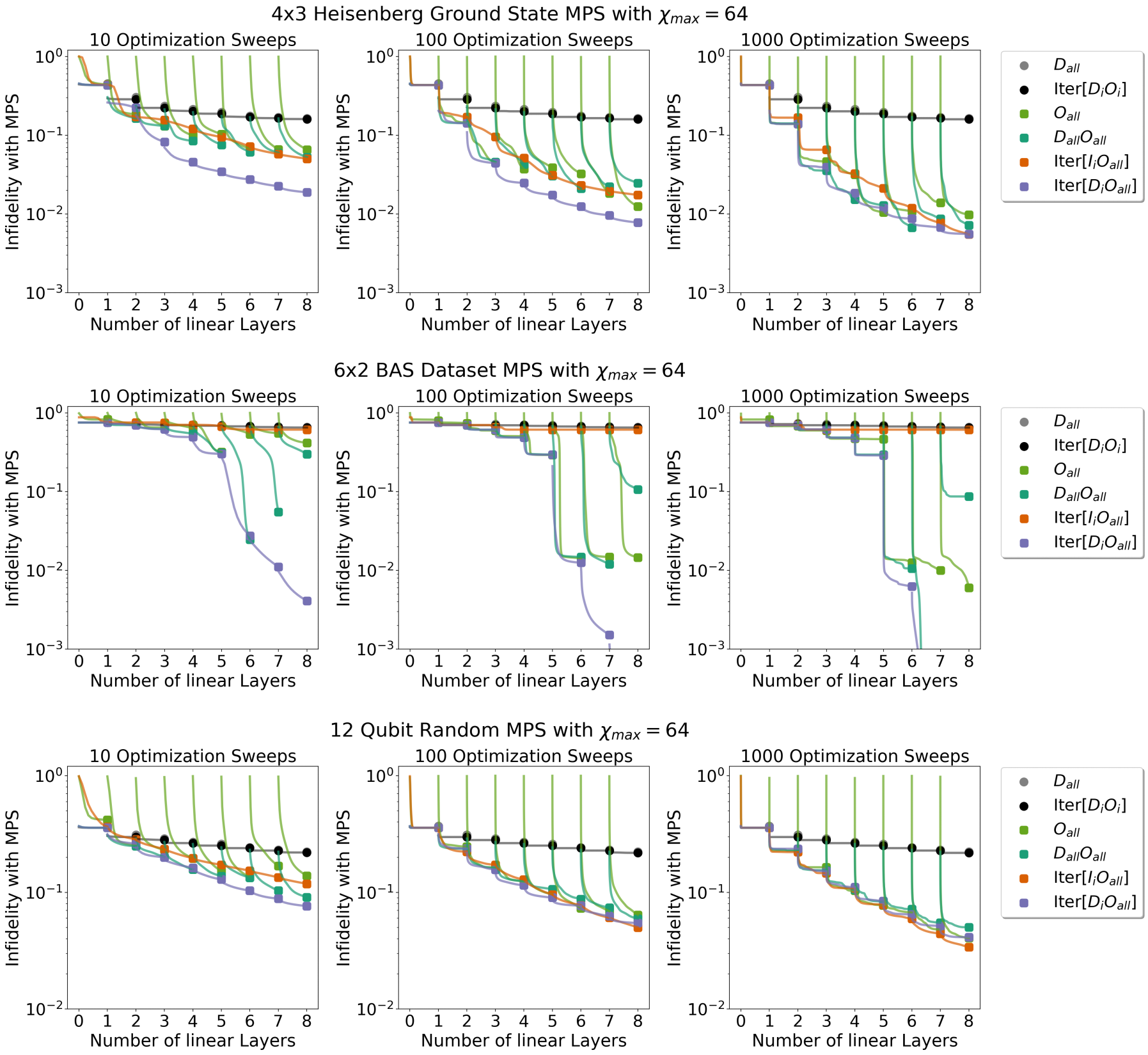}
	\caption{Benchmark results of the MPS decomposition protocols presented in Sec.~\ref{ssec:combination_decomposition} and Fig.~\ref{fig:decomposition_protocols}. We plot the infidelity of three target MPS with the respective quantum state prepared by the decomposed quantum circuit. We study the MPS examples which encode the ground state of a $4\times 3$ Heisenberg Hamiltonian (top), a uniform superposition over data samples from the 6x2 BAS dataset (middle), as well as a random MPS (bottom). All MPS have $\maxbd=64$. We report the infidelity with $T=10, T=100$ or $T=1000$ optimization sweeps per data point, as well as an illustration of the infidelity convergence during optimization (where applicable) in-between data points. 
	}
	\label{fig:decomposition_results}
\end{figure*}

In this Section, we quantitatively compare the six MPS decomposition protocols presented in Sec.~\ref{ssec:combination_decomposition}. We study three 12-qubit MPS with bond dimension $\maxbd = 64$ that represent the solutions to a Hamiltonian ground state search problem, a generative modelling task, as well as a random MPS.

The Hamiltonian is a spin $1/2$ antiferromagnetic Heisenberg model Hamiltonian with 
\begin{equation}\label{eq:heisenberg_hamiltonian}
	H =  \sum_{\langle i, j \rangle} \hat{S}_{X}^{(i)} \hat{S}_{X}^{(j)} + \hat{S}_{Y}^{(i)} \hat{S}_{Y}^{(j)} + \hat{S}_{Z}^{(i)} \hat{S}_{Z}^{(j)},
\end{equation}
where $\hat{S}_{X,Y,Z}=\frac{1}{2}\sigma_{x,y,z}$ are the spin operators acting on sites $i$ and $j$, and $\sigma_x, \sigma_y, \sigma_z$ are the Pauli operators. In particular, we study a $4\times3$ 2-dimensional Heisenberg model with open boundary condition, where $\langle i, j \rangle$ denotes the interaction of the respective vertical and horizontal neighbors in a grid layout. The MPS with $\maxbd=64$ exactly represents the ground state of this Hamiltonian. 

The second $\maxbd=64$ MPS exactly encodes a uniform superposition over the binary data samples in the $6\times2$ bars and stripes (BAS) dataset~\cite{Benedetti2019}. This dataset has emerged as a canonical benchmark for generative modelling tasks in the domain of quantum machine learning. The BAS MPS is special in that the singular value spectrum in the middle bond consists of either constant degenerated large values, or zero. As such, we expect results on this MPS to exhibit sharper increases in fidelity when all singular values can be reproduced by a sufficiently deep circuit.

Finally, we also test the decomposition protocols on a random $\maxbd=64$ MPS, where each with each tensor entry (before canonicalization) is drawn from a normal distribution with zero mean and variance of $1$. There are different conventions used in literature as to how to construct a random MPS state. We note that the existence of both negative and positive values presents a more challenging decomposition task, as the singular values decay significantly slower than with only positive tensor entries. 

Figure~\ref{fig:decomposition_results} depicts the decomposition performances on the three target MPS using the six protocols presented in Sec.~\ref{ssec:combination_decomposition}. We record the infidelity with the respective target MPS per number of circuit layers $K$, where infidelity is defined as $1-f$, with the fidelity $f$ from Eq.~\eqref{eq:fidelity_analytic} or equivalently Eq.~\eqref{eq:fidelity_optimization}. Note that some other works choose $1-f^2$. In addition to the finally attained infidelities, where applicable, we illustrate the optimization progress leading up to this final result in between the data points. This highlights how the sequential protocols (indicated by ``Iter'') build upon previous circuit layers. 
We test the performance of the decomposition protocols for $T=10, T=100$ or $T=1000$ optimization sweeps per number of layers. The cases $T=10$ and $T=100$ are more representative for results one may obtain in practice with a limited computational budget, whereas $T=1000$ highlights the robustness to local minima during optimization. Because the protocols Iter$[I_iO_{all}]$ and Iter$[D_iO_{all}]$ utilize the solution from the previous data point, and thus there are already optimization sweeps invested for any $k>1$, $T$ is increased in $O_{all}$ and $D_{all}O_{all}$ to match the total number of unitary optimization sweeps (compare also Eqs.~\eqref{eq:complexity_opt_plain} and~\eqref{eq:complexity_opt_sequential}). One important note is that we do not perform any truncation of singular values in these numerical results, but perform contractions with the exact statevector.

Our results show that the protocols \protocolone and \protocoltwo improve only very slowly with additional layers, which was also observed in Ref.~\cite{dov2022approxencoding}. In contrast, the protocol \protocolthree is mainly hindered by the large number of optimization sweeps $T$ necessary to achieve high-quality approximations, but as we see when comparing $T=100$ and $T=1000$, the protocol may additionally run into complications related to local minima. While local minima for larger $K$ also occur for the protocol \protocolfour, it achieves significantly improved approximations for low $T$. In the cases of the ground state and random MPS, where the singular value spectra are more continuous, the protocol \protocolfive can achieve competitive performance. However, for the BAS dataset MPS, the protocol is clearly under-performing. Finally, it becomes evident that the protocol \protocolsix is the best-performing and most consistent across the all target MPS and number of optimization sweeps. It is less affected by sub-optimal local minima than \protocolthree and \protocolfour, and can achieve signficantly better fidelities than \protocolfive on the BAS dataset MPS. The improvements are most significant for $T=10$ and $T=100$, which is a very desirable property that reduces the amount of classical computing resources spent for a given decomposition fidelity. 

To summarize, we find the protocol \protocolsix to be the most performant and promising for practical application. When only a relatively small amount of resources are available for optimization, we find the improvement in infidelity per number of layers $K$ reach orders of magnitude relative to other protocols. In particular, our ablation analysis shows that, if one removes any one component of the protocol, namely the analytical decomposition, the optimization, or the sequential growing scheme of the circuit, the protocol performs significantly worse.

\section{Conclusion}
In this work, we presented a range of algorithmic protocols to decompose MPS with arbitrary bond dimension into linear quantum circuit layers consisting of two-qubit unitaries. The fundamental building-block algorithms for the protocols are the analytical decomposition algorithm presented in Ref.~\cite{ran2020encoding}, and a constrained optimization algorithm to optimize the circuit unitaries on classical hardware, which was utilized in Ref.~\cite{shirakawa2021automatic} (see also the earlier method of Ref.~\cite{evenbly2009algorithms}). We benchmarked the MPS decomposition protocols on three different 12-qubit MPS with bond dimension $\chi=64$. 
Across our benchmarks, we found the protocol \protocolsix to be the most successful in reliably decomposing MPS with high fidelity into shallow quantum circuit layers. The protocol combines sequential growing the quantum circuit using the analytical decomposition algorithm with intertwined optimization steps of all existing circuit unitaries. It performs particularly well with a limited number of optimization steps - sometimes achieving orders of magnitude improvement in infidelity compared to other protocols. Nevertheless, further quantitative studies are necessary to evaluate the performance and scalability of this MPS decomposition protocol.

A promising follow-up work is to extend the protocols presented here to TTNs. TTNs are more flexible than MPS in terms of the long-range correlations that they can capture, a feature which has already proven practically beneficial in generative modeling tasks~\cite{Cheng2019}. By efficiently converting TTNs into PQCs running on quantum hardware, these benefits can likely be translated into better performance on a variety of simulation and quantum machine learning tasks.
Decomposing the two-dimensional PEPS on classical hardware may be significantly more challenging, due to the lack of strong canonical forms facilitating the exact conversion into circuits of multi-qubit unitaries.
However, the close resemblance between the graphical structure of PEPS and that of physical qubit topologies in typical quantum hardware (e.g.~\cite{Google2019supremacy}) could make such conversions from PEPSs to PQCs extremely promising for practical applications.
While our driving motivation in this work has been to limit the amount of quantum resources spent for the TN decomposition, one may attempt to perform certain sub-routines on quantum hardware~\cite{dov2022approxencoding}. This could be particularly useful when working with PEPS~\cite{Ian_PEPS_to_QC,Lei_VQE_PEPS}, or when aiming to map MPS or TTNs to circuit layouts that are not classically efficient to simulate.

An accompanying work, Ref.~\cite{rudolph2022synergistic}, utilizes the MPS decomposition protocol presented here inside a synergistic optimization framework, and demonstrates how the performance of PQCs is continuously improved by MPS with increasing $\maxbd$ across several applications. Additionally, Ref.~\cite{rudolph2022synergistic} provides evidence that PQCs initialized with classically trained TNs may not be affected by barren plateaus~\cite{Mcclear2018Barren,cerezo2021costfunction,holmes2022expressivity}, due to the highly specific circuit initialization. This companion work can be seen as the realization of the benefits predicted in Ref.~\cite{Huggins_2019}, which uses our decomposition protocol to convert MPS into shallow quantum circuits. We are optimistic that future work will identify yet greater improvements on our methods, and foster a rich interconnection between the growing TN and PQC research communities.




%



\appendix

\section{Unitaries \& Layers - Notation and Convention}\label{apx:conventions}
In this work, we use the notation $U$ for general \ufour unitaries acting on two qubits. A collection of unitaries $\{\unitary{m}\}_m$ will be indexed by the index $m$. 

Linear layers of unitaries arranged in a \textit{stair-case} or \textit{next-neighbor} topology (see for example the layout in Figs.~\ref{fig:mps_mapping}\&\ref{fig:fundamental_algorithms}) is denoted as L$[U]$. A collection of layers $\{ \layer{k}\}_k$ will be indexed by the index $k$.

Whenever unitaries or layers of unitaries are applied to a state $\psi$, we will be indexing the order according to the application to the \textit{ket} state $\ket{\psi}$. 

In particular, in Sec.~\ref{ssec:optimization} we apply unitaries via 
\begin{equation}
	\begin{aligned}
		\ket{\phi} &= \prod_{m=1}^M \unitary{m} \ket{\psi}\\
		&= \unitary{M}\dots \unitary{2}\unitary{1} \ket{\psi},
	\end{aligned}
\end{equation}
and analogous
\begin{equation}
	\begin{aligned}
		\bra{\phi} &= \bra{\psi} \left(\prod_{m=1}^M \unitary{m}\right)^\dagger \\
		&= \bra{\psi} \prod_{m=M}^1 \unitary{m}^\dagger \\
		&= \bra{\psi} \unitary{1}^\dagger \dots \unitary{M}^\dagger.
	\end{aligned}
\end{equation}

In the case of layers of unitaries, like in Sec.~\ref{ssec:analytic}, we follow
\begin{equation}
	\begin{aligned}
		\ket{\phi} &= \prod_{k=1}^K \layer{k} \ket{\psi}\\
		&= \layer{K}\dots \layer{1} \ket{\psi},
	\end{aligned}
\end{equation}
and analogous
\begin{equation}
	\begin{aligned}
		\bra{\phi} &= \bra{\psi} \left(\prod_{k=1}^K \layer{k}\right)^\dagger \\
		&= \bra{\psi} \prod_{k=K}^1 \layerdagger{k} \\
		&= \bra{\psi} \layerdagger{1} \dots \layerdagger{K},
	\end{aligned}
\end{equation}
where the order of unitaries in side each layer is reversed, and all unitaries are conjugated and transposed.

\clearpage

\end{document}